\documentclass[%
prl,
 reprint,
superscriptaddress,
 amsmath,amssymb,
 aps,
floatfix,
]{revtex4-2}

\usepackage{amsmath}
\usepackage{amssymb}
\usepackage{physics}
\usepackage{siunitx} \AtBeginDocument{\RenewCommandCopy\qty\SI}
\usepackage[colorlinks=true,
            linkcolor=red,
            citecolor=blue,
            urlcolor=blue]{hyperref}
\usepackage{resizegather}
\usepackage{graphicx}
\usepackage{booktabs} 
\usepackage{csquotes} 
\renewcommand{\selectlanguage}[1]{}
\DeclareUnicodeCharacter{03B2}{{$\beta$}}
\DeclareUnicodeCharacter{2207}{{$\nabla$}}

\begin{document}

    \title{The pellet rocket effect in magnetic confinement fusion plasmas}

    \author{Nico J. Guth}
\affiliation{Department of Physics, Chalmers University of Technology,
  SE-41296 Gothenburg, Sweden}
\affiliation{Max Planck Institute for Plasma Physics, 17491 Greifswald, Germany}
\author{Oskar Vallhagen}
 \affiliation{Department of Physics, Chalmers University of Technology,
   SE-41296 Gothenburg, Sweden}
 \author{Per Helander}
 \affiliation{Max Planck Institute for Plasma Physics, 17491 Greifswald, Germany}
\author{Istvan Pusztai}
\affiliation{Department of Physics, Chalmers University of Technology,
  SE-41296 Gothenburg, Sweden}
  \author{Sarah L. Newton}
 \affiliation{United Kingdom Atomic Energy Authority, Culham Campus, Abingdon, Oxon OX14 3DB, UK}
 
 \author{T\"unde F\"ul\"op}
\affiliation{Department of Physics, Chalmers University of Technology,
  SE-41296 Gothenburg, Sweden}
  \affiliation{Rudolf Peierls Centre for Theoretical Physics, University of Oxford, Oxford, OX1 3PU, UK}
\affiliation{Merton College, Oxford, OX1 4JD, UK}
    \begin{abstract}
Pellets of frozen material travelling into a magnetically confined fusion plasma are accelerated by the so-called \emph{pellet rocket effect}.
The non-uniform plasma heats the pellet ablation cloud asymmetrically, producing pressure-driven, rocket-like propulsion of the pellet. 
We present a semi-analytical model of this process by perturbing a spherically symmetric ablation model.
Predicted pellet accelerations match experimental estimates in current tokamaks ($\sim 10^5 \; \rm{m/s^2}$). Projections for ITER high-confinement scenarios ($\sim 10^6 \; \rm{m/s^2}$) indicate significantly shorter pellet penetration than expected without this effect, which could limit the effectiveness of disruption mitigation.
    \end{abstract}
\maketitle    

Injection of pellets of frozen material is a primary actuator for density profile control in magnetic confinement fusion (MCF) plasmas. It is regularly used for refuelling, controlling edge localized modes, and mitigating disruptions \cite{pegourie_review_2007}. Although pellet modelling often assumes uniform linear pellet motion, experimental observations suggest that pellets injected into MCF plasmas are toroidally deflected and significantly accelerated towards the low-field side, which affects the fuelling efficiency and material deposition profile \cite{kong_interpretative_2024}.
While the complete physics of this pellet acceleration is not fully understood, it is commonly attributed to the \emph{pellet rocket effect}.

Asymmetries in the heat flux onto the pellet surface enhance the ablation on one side of the pellet and lead to asymmetric heating of the pellet ablation cloud.
Consequently, the pellet is pushed, similarly to a rocket, in the opposite direction to the ejected material,  
modifying the pellet's trajectory \cite{kong_interpretative_2024,jachmich_shattered_2022}.
Even though several studies indicate that the pellet rocket effect might become significant in reactor-scale devices \cite{samulyak_simulation_2023,kamelander_modelling_2008}, it is still mostly neglected in state-of-the-art pellet modelling, and computationally inexpensive models from first principles are lacking. 

The material deposition profile hinges on both the pellet trajectory through the plasma and the rate of ablation.
The ablation rate of pellets has been studied extensively \cite{pegourie_review_2007}. 
Good agreement with experiments \cite{houlberg_neutral_1988,houlberg_pellet_1992,rozhansky_ablation_2005,cseh_pellet_2017} has been achieved by one of the earliest ablation models, the neutral gas shielding (NGS) model by \textcite{parks_effect_1978}. 
In this model, the energy flux from the plasma onto the pellet is taken to be equal to that of a mono-energetic beam of hot electrons.
The resulting ablated material is assumed to form a dense neutral gas cloud that undergoes transonic radial expansion and shields the pellet from most of the energy flux of incoming plasma electrons (and ions). 
More sophisticated models are not found to yield significantly higher accuracy \cite{pegourie_review_2007}, and the NGS model is widely used for its low computational complexity.

There have been a limited number of theoretical studies of the pellet rocket effect. The study by \textcite{andersen_injection_1985} connected toroidal pellet deflection to the asymmetry stemming from the plasma current. 
\textcite{senichenkov_pellet_2007} assumed the ablation cloud to be homogeneous and attributed the acceleration purely to an enhanced ablation rate on one side of the pellet.
\textcite{szepesi_radial_2007} proposed a semi-empirical model, in which the pressure asymmetry is the driving factor of the acceleration but must be given as a model parameter in addition to the NGS model scaling laws.

Here, we develop a semi-analytical theory of the rocket effect for hydrogenic pellets, with the asymmetry self-consistently calculated. 
The basis of our model is the NGS model with the asymmetry treated as a perturbation.
This leads to a linear connection between the asymmetric heating of the neutral cloud and the pressure asymmetry at the pellet surface, which can be used to compute the pellet rocket acceleration.  

Consider a spherical pellet of radius $r_\text{p}$ surrounded by ablated material in the form of a charge-neutral gas.
As we shall see below, the ablated material travels radially outward with a velocity that increases monotonically with radius and reaches the sound velocity at a certain radius. Mathematically, the problem is similar to the one of the solar-wind, where the flow velocity is found by solving the fluid equations with gravity. Physically, the force on the pellet arises from the combination of ablated particles leaving the pellet surface and the gas pressure pushing on the pellet surface.

\begin{figure}[htpb]
 \includegraphics[width=\linewidth]{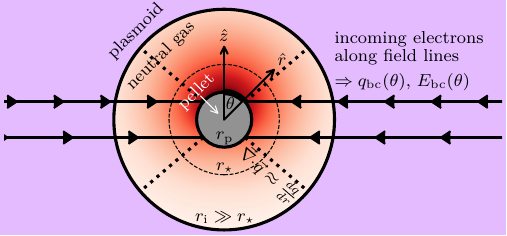}\vspace{-0.2cm}
  \caption{A pellet (grey) and its surrounding ablation cloud (neutral gas in red, ionized in purple) are heated by hot electrons from the background plasma.
The ablated material undergoes a near-spherical, accelerated outflow, surpassing the speed of sound at radius $r_\star$ and becoming ionized much further away at radius $r_\text{i} \gg r_\star$. The electron heat flux, incoming along the magnetic field lines, increases in the $\hat{z}$ direction.  For example, $\hat{z}$ could be the direction of the background plasma temperature gradient. The electron path through the neutral cloud will be approximated by an equivalent radial path (dotted line).}
  \label{fig:neutralcloud}
\end{figure}

To calculate this force, we choose spherical coordinates $\{r,\theta,\varphi\}$, as illustrated in Fig.~\ref{fig:neutralcloud}, so that $r$ is the radial distance from the pellet centre and $\varphi$ is the azimuthal angle. The polar axis is aligned with $\hat{z}$, the direction along which the external heat flux, incoming along magnetic field lines, increases.
The rocket force is defined along the $-\hat{z}$-direction, that is, a positive force pushes the pellet away from the region of higher heat flux.
The force can be written as an integral over the pellet surface, $S$,
\begin{gather*}
    F = - \hat{z} \cdot \vec{F} = r_\text{p}^2 \iint_{S} \left[ \rho v_r (v_r \cos\theta + v_\theta \sin\theta) + p \cos\theta \right] \,\dd{\Omega},
\end{gather*} 
with differential solid angle $\dd{\Omega}=\sin\theta \dd{\theta}\dd{\varphi}$, where the neutral gas is described by its flow velocity $\vec{v} = v_r \hat{r} + v_\theta \hat{\theta} + v_\varphi \hat{\varphi}$, pressure $p$ and mass density $\rho$. 
The asymmetry is taken as a perturbation to the otherwise spherically symmetric dynamics. 
Expanding the perturbed variables of the neutral cloud in the Legendre polynomials, $P_l(\cos \theta)$, allows computation of $F$ using only the $l=1$ component.
That is, without loss of generality, we may write a quantity $y\in \{\rho, p, T, v, q, E\}$ as 
\begin{equation}
    y(r,\theta) = y_0(r)+y_1(r)\cos \theta,
    \label{eq:y_expansion}
\end{equation} 
where $T=m p/\rho$ is the neutral cloud temperature in units of energy, with $m$ the mass of the neutral particles, $q$ is the incident electron energy flux, and $E$ is the average energy of those electrons.
The only exception is the flow velocity, which we take as $\vec{v} = \hat{r} [v_0(r)+v_{1,r}(r) \cos\theta] - \hat{\theta} v_{1,\theta}(r) \sin\theta$. 
With these definitions, the pellet rocket force becomes 
\begin{equation}
    F = (4 \pi r_\text{p}^2/3) \left[ \rho_1 v_0^2 + 2 \rho_0 v_0 (v_{1,r} -v_{1,\theta}) + p_1 \right]_{r=r_\text{p}}.
    \label{eq:rocket_force_full}
\end{equation}
The first two terms in Eq.~(\ref{eq:rocket_force_full}) describe the force arising from asymmetric ablation.
The term $\rho_0 v_0 v_{1,\theta}$ describes a force from mass flowing around the pellet surface, and the last term $p_1$ describes the gas pressure asymmetry. 

Based on the spherically symmetric NGS model for $y_0$, a set of equations determining the $r$-dependence of the asymmetric perturbation variables $y_1$ can be derived from the ideal gas law and steady state mass, momentum and energy conservation as 
\begin{align}
    & T_0^2 \rho_1 = m \left( p_1 T_0 - p_0 T_1 \right) \, , \label{eq:physical_perturbation_ideal_gas_law} \\
    & r^2 v_{1,r} \partial_r\rho_0 + \rho_0 \left[ \partial_r (r^2 v_{1,r}) - 2r v_{1,\theta} \right]\nonumber\\ 
    &+ r^2 v_0 \partial_r\rho_1 + \partial_r(r^2 v_0) \rho_1 = 0 \, , \label{eq:physical_perturbation_mass_conservation} \\
    & \rho_0 v_0 \partial_r v_{1,r} + (\rho_0 v_{1,r} + v_0 \rho_1) \partial_r v_0 = - \partial_r p_1 \, ,\label{eq:physical_perturbation_r_momentum_conservation}\\
    & r \rho_0 v_0 \partial_r v_{1,\theta} + \rho_0 v_0 v_{1,\theta} = - p_1 \, , \label{eq:physical_perturbation_theta_momentum_conservation} \\
    & \left[ r^2 v_{1,r} \partial_r + \partial_r(r^2 v_{1,r}) - 2r v_{1,\theta} \right] \left( 0.5\rho_0 v_0^2  + \Upsilon p_0  \right) \nonumber \\
    & + \left[ r^2 v_0 \partial_r + \partial_r(r^2 v_0) \right] \left( 0.5 \rho_1 v_0^2  + \rho_0 v_0 v_{1,r} + \Upsilon p_1 \right) \nonumber \\
    & = Q \partial_r q_1\, ,\label{eq:physical_perturbation_energy_conservation}
\end{align}
where $\Upsilon = \gamma / (\gamma -1)$, and $\gamma$ denotes the adiabatic index of the gas.
The heating source on the right-hand side of Eq.~\ref{eq:physical_perturbation_energy_conservation} is taken to be a fraction $Q$ of the local loss of electron energy flux $-\vec{\nabla}\cdot \vec{q} \approx \partial_r q$ along an equivalent radial path of the electrons (see Fig.~\ref{fig:neutralcloud}). $Q$ is estimated to be $60$-$70\%$ \cite{parks_model_1977}.
(The remainder is mainly lost through Bremsstrahlung radiation and backscattered electrons.) 

The radial variation of the electron energy flux and the average electron energy are described by
\begin{align}
     &m \partial_r q_1 = (\rho_1 q_0  + \rho_0 q_1) \Lambda(E_0) + \rho_0 q_0 E_1 \left.{\partial_E \Lambda}\right|_{E=E_0} \, , \label{eq:physical_perturbation_effective_heat_flux} \\
    &m \partial_r E_1 = 2 \rho_1 L(E_0) + 2 \rho_0 E_1 \left.{\partial_E L}\right|_{E=E_0} \, , \label{eq:physical_perturbation_electron_energy_loss}   
\end{align}
with the effective energy flux cross-section $\Lambda(E) = \hat{\sigma}_T(E) + 2 L(E)/E$. 
We use the empirical energy loss function, $L(E)$, derived by \textcite{miles_electron-impact_1972} and the effective backscattering cross-section, $\hat{\sigma}_T(E)$, given by \textcite{parks_model_1977-1} derived from experimentally measured values by \textcite{maecker_ionen-_1955}. 
These expressions are valid for electron scattering in a molecular hydrogen gas.

The low sublimation energy of hydrogen implies that the pellet is almost fully shielded by its ablation cloud, which is reflected by the imposed boundary condition $q(r_{\rm p})=0=T(r_{\rm p})$. Additionally, we assume $v_\theta(r_{\rm p})=0$ and $p(r\rightarrow \infty)=0$.
A convenient way to quantify the degree of asymmetry in the external heat source is by defining $q_\text{rel} = q_{{\rm bc} 1}/q_{{\rm bc} 0}$ and $E_\text{rel} = E_{{\rm bc} 1}/E_{{\rm bc} 0}$, where the quantities $q_{{\rm bc}}$ and $E_{{\rm bc}}$ are the angularly dependent heating boundary conditions for $r \rightarrow \infty$, expanded according to Eq.~(\ref{eq:y_expansion}).
The perturbation assumption requires $|q_\text{rel}| \ll 1$ and $|E_\text{rel}| \ll 1$.

Before solving Eqs.~(\ref{eq:physical_perturbation_ideal_gas_law}) to (\ref{eq:physical_perturbation_electron_energy_loss}) numerically under these boundary conditions, all quantities are conveniently normalized to reduce the number of free parameters.
Spherically symmetric quantities $y_0$ are normalised by their values, $y_\star=y(r_\star)$, at the radial location $r_\star$ where the flow velocity reaches the sound speed. The perturbation quantities $y_1$ are also normalised to those values multiplied by $q_\text{rel}$. The normalized solutions are then fully determined by providing the adiabatic index $\gamma$, the spherically symmetric component of the incoming electron energy $E_{{\rm bc0}}$ and the degree of asymmetry in the external heat source as $E_\text{rel}/q_\text{rel}$. The physical solution can later be inferred by additionally providing $r_\text{p}$, $q_\text{bc0}$ and $q_\text{rel}$.

\begin{figure}[htpb]
    \centering
    \includegraphics{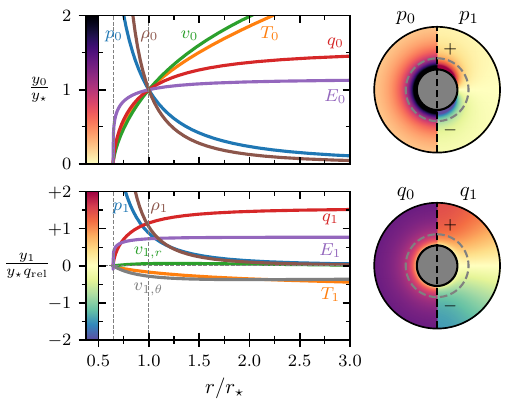}\vspace{-0.2cm}
    \caption{Spatial dependence of the perturbative ablation dynamics. The line plots on the left show the radial dependence of the NGS solution (top) and the perturbation solution (bottom). To the right, the full spatial dependence is visualized for the pressure (top) and electron energy flux (bottom) for both the NGS model (left half) and perturbation (right half). The dashed circle marks the sonic radius. The parameters are $\gamma=7/5$, $E_\text{bc0}=\qty{4}{\kilo\eV}$, $E_\text{rel}/q_\text{rel}=2/3$.}
    \label{fig:results1d}
\end{figure}

Fig.~\ref{fig:results1d} illustrates the obtained ablation cloud behaviour, with details of the radial profiles of all variables in the cloud, as well as overviews of the 2D structure of the pressure $p$ and electron energy flux $q$. 
The cross-sections show the 2D spatial variation of the spherically symmetric dynamics (on the left) and the asymmetric perturbation (on the right).
The quantities are presented through a variation in colour, where darker values mean higher absolute values. 
For the perturbation quantities, red denotes an increase and blue denotes a decrease compared to the NGS model.
Note that the figure shows all quantities as their normalized version -- the physical perturbations are much smaller than their spherically symmetric values.
The pellet is visualized by the grey circle in the middle, and the dashed line around it indicates the sonic radius $r_\star$.
Only the region close to the pellet is shown, and ionization occurs at much larger radii.
The heating of the neutral gas from the incoming energy flux mostly takes place close to the pellet, where $q$ and $E$ drop rapidly.
Interestingly, the temperature perturbation $T_1$ is negative, which means that the temperature is higher on the side opposite to the stronger heating.
This can be understood by noticing that the enhanced ablation produces a positive density perturbation $\rho_1$, which increases the thermal capacity of the ablation cloud.

Fig.~\ref{fig:results1d} shows one specific normalized solution ($\gamma=7/5$, $E_\text{bc0}=\qty{4}{\kilo\eV}$, $E_\text{rel}/q_\text{rel}=2/3$), which is representative of the example tokamak predictions presented at the end of this paper.
Quantitative analysis of many solutions over physically interesting ranges of the parameters $\gamma$, $E_\text{{\rm bc}0}$ and $E_\text{rel}/q_\text{rel}$ \cite{guth_pellet_2024-1} has shown that the pressure asymmetry $p_1(r_\text{p})$ is the dominant term in Eq.~(\ref{eq:rocket_force_full}).
Adding the other terms changes the rocket force only in the third or fourth significant figure, i.e.
\begin{equation}
    F \approx (4\pi r_\text{p}^2/3) p_1(r_\text{p}) \, .
    \label{eq:force_with_only_pressure}
\end{equation}

\begin{figure}[htpb]
    \centering
    \includegraphics{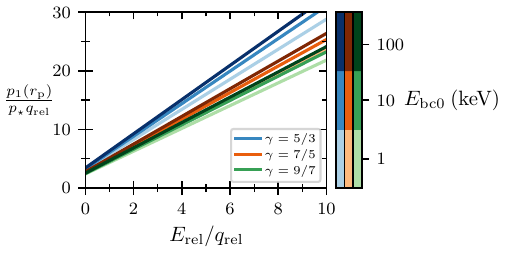}\vspace{-0.2cm}
    \caption{Linear dependence of the normalized pressure asymmetry at the pellet surface on the parameter $E_\text{rel}/q_\text{rel}$. The colour of the lines denote $\gamma$, while the brightness denotes $E_\text{bc0}$.}
    \label{fig:P1_at_r_p}
\end{figure}

\begin{figure}
    \centering
    \includegraphics{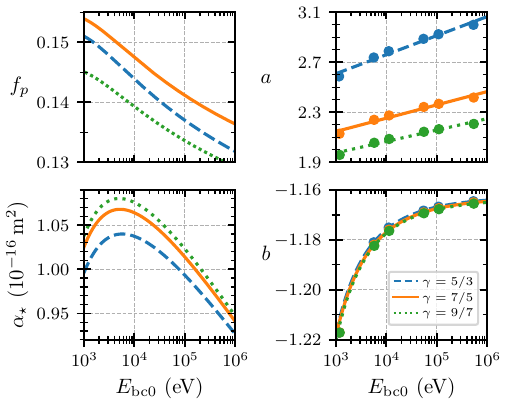}
    \vspace{-0.2cm} 
    \caption{Weak energy dependence of the quantities $f_p$, $a$, $b$ and $\alpha_\star$. $f_p$ and $\alpha_\star$ in Eq.~(\ref{eq:pstar}) are calculated by reproducing NGS model solutions \cite{parks_effect_1978}. $a$ and $b$ in Eq.~(\ref{eq:p1_at_r_p}) are calculated through linear regression on $p_1(r_\text{p})/p_\star q_\text{rel}$ scanned over $E_\text{rel}/q_\text{rel}$ for six values of $E_\text{bc0}$ and three values of $\gamma$ (marked by dots), as shown in Fig.~\ref{fig:P1_at_r_p}. The lines for $a$ and $b$ represent polynomial fits.}
    \label{fig:parameters}
\end{figure}

Furthermore, the qualitative dynamics barely change with $\gamma$ and $E_\text{{\rm bc}0}$, but the dependence on $E_\text{rel}/q_\text{rel}$ is mostly linear.
This is illustrated for the relevant quantity $p_1(r_\text{p})/p_\star q_\text{rel}$ in Fig.~\ref{fig:P1_at_r_p}.
We find the simple relationship
\begin{equation}
    p_1(r_\text{p}) = a p_\star  \left(E_\text{rel} - b q_\text{rel}\right) \, ,
    \label{eq:p1_at_r_p}
\end{equation}
where the linear regression parameters $a \approx 2.0$ to $2.9$ and $b \approx -1.21$ to $-1.17$ depend weakly on $\gamma$ and $E_\text{bc0}$, as shown in Fig.~\ref{fig:parameters}.
The pressure at the sonic radius is given by the NGS model as
\begin{gather}
    p_\star = \underbrace{ \frac{\lambda_\star}{\gamma} \left( \frac{(r_\text{p}/r_\star) (\gamma-1)^2}{4 (q_{{\rm bc}0}/q_\star)^2} \right)^\frac{1}{3} }_{f_p(E_{\text{bc}0}, \gamma)} \left[ \frac{m (Q q_{\text{bc}0})^2}{\alpha_\star \, r_\text{p}} \right]^\frac{1}{3} \left(\frac{E_\text{bc0}}{\unit{\eV}} \right)^\frac{1.7}{3} \, ,
    \label{eq:pstar}
\end{gather}
with the definitions $\lambda_\star=\rho_\star r_\star \Lambda(E_{\star})/m$ and $\alpha_\star = (E_\text{bc0}/\unit{\eV})^{1.7} \Lambda(E_{\star})$ \cite{parks_effect_1978,guth_pellet_2024-1}. 
It was already recognised (in a slightly different form) by \textcite{parks_effect_1978} that the quantities $f_p\approx 0.15$ and $\alpha_\star \approx \qty{1.1e-16}{\m^2}$ are nearly constant over the $E_\text{bc0}$ range of interest, as shown in Fig.~\ref{fig:parameters}, making it possible to derive fairly simple scaling laws. 
Overall, Eqs.~(\ref{eq:force_with_only_pressure}), (\ref{eq:p1_at_r_p}) and (\ref{eq:pstar}) in combination with the values shown in Fig.~(\ref{fig:parameters}) suffice for predicting the pellet rocket force for a given heating source ($E_\text{bc0}$, $E_\text{rel}$, $q_\text{bc0}$, $q_\text{rel}$) and pellet radius $r_\text{p}$.
The analysis assumed a spherical pellet, but the results found here also apply to pellets with small deviations from spherical symmetry: upon expanding the pellet surface shape and the hydrodynamic quantities in spherical harmonics, the net force on the pellet, Eq.~(\ref{eq:rocket_force_full}), acquires an additional contribution that is linear in the $l=1$, $m=0$ harmonic of its surface shape. As this component corresponds to a uniform shift of the surface, the corresponding net force must vanish.    

The asymmetry in the incoming energy flux that produces the rocket force originates from the fact that the plasma surrounding the pellet is non-uniform. Not only are there gradients in the background plasma density and temperature, but in addition, the curvature of the magnetic field causes newly ionised ablated material to drift across field lines and create a non-uniform shielding plasmoid distribution around the neutral cloud \cite{lang_high-efficiency_1997,parks_radial_2000,vallhagen_drift_2023}. 
The latter effect was considered by \textcite{senichenkov_pellet_2007} as the primary source of heat flux asymmetry, enhancing the ablation on one side of the pellet. 
We find that even the small asymmetry due to the background temperature and density gradients across the pellet is sufficient to produce significant acceleration. This can be understood by noticing that the pressure at the pellet surface reaches around $\qty{100}{atm}$, of which a small relative perturbation can lead to large enough absolute force values.

As a first demonstration of our model, we consider only this effect of asymmetry induced by the background plasma electron density, $n_\text{bg}$, and temperature, $T_{\text{bg}}$.
Adding the effect of the drifting plasmoid will further increase the magnitude of the rocket acceleration, which will be discussed in a future study \cite{guth_effect_2024}.
In a typical MCF plasma, the gradients are perpendicular to the magnetic field lines.
The corresponding rocket force points in the direction of decreasing temperature and density, i.e. outwards along the minor radius of a tokamak.
Referring to Fig.~\ref{fig:neutralcloud}, we thus let the gradients point in the positive $\hat{z}$-direction, with $z=0$ marking the field line passing through the pellet centre. 
The local effective electron energy and energy flux can be expressed in terms of the background plasma parameters as
    $E_{\text{bc0}} = 2 T_{\text{bg}}$ and $q_{\text{bc0}} = 2 n_{\text{bg}}\sqrt{T^3_{\text{bg}}/(2 \pi m_e)} $ \cite{parks_model_1977}. 
To first order, the heating asymmetry without plasmoid shielding then becomes 
\begin{align}
   E_\text{rel} &= \delta r \left[ (\partial_z T_\text{bg}) / T_\text{bg} \right] \label{eq:E_rel} \quad \text{and} \\
   q_\text{rel} &= \delta r \left[ 3 (\partial_z T_\text{bg})/2 T_\text{bg} +  (\partial_z n_\text{bg}) / n_\text{bg} \right] \, , \label{eq:q_rel}
\end{align}
where $\delta r$ is an effective radius of the heat deposition, defined by the variational parameter $\delta z = \delta r \cos\theta$. 
The dominant heating close to the pellet (within the sonic radius $r_\star \approx 1.5 r_\text{p}$ \cite{parks_effect_1978}), combined with the radial electron path mapping (see Fig.~\ref{fig:neutralcloud}), suggests $r_\text{p} \leq \delta r \leq 1.5 r_\text{p}$. 

We are now in a position to apply our self-consistent semi-analytical model for the pellet rocket force to example cases.
The injected pellets are assumed to consist of pure deuterium. 
This makes the neutral cloud around the pellet a mostly diatomic gas of ${D_2}$ molecules, which suggests the adiabatic index $\gamma = 7/5$ and the gas particle mass $m = \qty{4}{u}$.
Furthermore, we choose $Q = 0.65$ and $\delta r = r_\text{p}$.
Assuming spherical pellets, the pellet rocket force, $F$, leads to an expression for the pellet rocket acceleration as
   $ \dot{v} = {F}/\left(\frac{4}{3} \pi r_\text{p}^3 \rho_\text{p}\right)$, 
where $\rho_\text{p} \approx {204}\; \rm kg/m^3$ \cite{senichenkov_pellet_2007} is the typical density of cryogenic deuterium pellets. 

\begin{table}[htbp]
    \centering
    \begin{tabular}{ccccc c}
        \toprule
        {$r_\text{p}$} & {$T_\text{bg}$} & {$\partial_z T_\text{bg}$} & {$n_\text{bg}$} & {$\partial_z n_\text{bg}$} & {$\dot{v}$} \\        
        {$(\unit{mm})$} & {$(\unit{\keV})$} & {$(\unit{\keV/\m})$} & {$(\unit{10^{19}/\m^{3}})$} & {$(\unit{10^{19}/\m^{4}})$} & {$(\unit{10^5 \m/\s^{2}})$} \\
       \midrule
1.0 & 0.3 & 10.0 & 3.0 & 80.0 & 1.0 \\
1.0 & 1.5 & 3.0 & 5.0 & 0.0 & 0.7 \\
0.5 & 2.0 & 5.0 & 9.0 & 0.0 & 2.7 \\ 
       \bottomrule
    \end{tabular}%
    \caption{Predictions of the pellet rocket acceleration $\dot{v}$, for example parameter combinations representative of AUG.}
    \label{tab:example_accelerations}
\end{table}

For given pellet radii 
and various local background plasma parameter combinations representative of the {ASDEX} Upgrade (AUG) tokamak \cite{meyer_overview_2019},  
the predicted rocket deceleration 
values are listed in Table \ref{tab:example_accelerations}.
The examples include both pedestal relevant large gradients and plasma core relevant values.
Pellet injection experiments by Müller \textit{et al.} \cite{muller_high-_1999,muller_high_2002} in AUG give measured average rocket acceleration values of $\sim \qty{4e5}{\m/\s^2}$.
Our predictions are of the same order of magnitude, but somewhat lower, likely resulting from the 
neglect of plasmoid shielding effects \cite{guth_pellet_2024-1,guth_effect_2024}.

To provide estimates for the impact of the rocket effect on pellet penetration on ITER, we numerically calculate simplified pellet trajectories through prescribed background plasma parameter profiles.
The pellet is assumed to be injected horizontally from the low-field side and travels into the plasma along a straight line.
In this calculation, we include the pellet ablation according to the NGS model \cite{parks_effect_1978} and the deceleration according to our model.
Any other dynamics, such as the influence of the ablated material on the background plasma or the plasmoid shielding, are neglected for simplicity.

We consider two commonly used ITER scenarios
produced by the CORSICA workflow \cite{kim_investigation_2018,vallhagen_runaway_2024}:
DTHmode24, a high-confinement mode (H-mode) D-T plasma with a core temperature of $\qty{22.6}{\keV}$ and electron density of $\qty{8.3e19}{\m^{-3}}$; and
H26, a low-confinement mode (L-mode) hydrogen plasma with a core temperature of $\qty{5.1}{\keV}$ and electron density of $\qty{5.2e19}{m^{-3}}$.
ITER pellets will cover a range of  sizes $\sim \qtyrange{1}{10}{\mm}$ and injection speeds $\sim \qtyrange{100}{1000}{\m/\s}$.
As a test case, we estimate how far a spherical pellet of $\qty{3}{\mm}$ radius injected at $\qty{500}{\m/\s}$ can travel into the plasma.
At constant velocity, 
such a pellet would be fully ablated after $\qty{76}{\cm}$ in the L-mode scenario and after $\qty{12}{\cm}$ in the H-mode scenario. 
Including the rocket effect, our model predicts a time-averaged deceleration of only $\qty{6.2e4}{\m/\s^2}$ in the L-mode scenario, while $\qty{2.7e6}{\m/\s^2}$ in the H-mode scenario.
The L-mode pellet trajectory would be barely affected ($\qty{3}{\cm}$ shorter).
However, the pellet in the H-mode scenario would be reflected before being fully disintegrated and would reach only $\qty{5}{\cm}$ into the plasma, which is $47\%$ of the penetration depth without deceleration.
Other test cases indicate that larger or slower pellets are generally even more affected by the rocket effect, while smaller or faster pellets are less affected.

A more accurate analysis of the rocket effect, including plasmoid shielding effects on the heating asymmetry of the ablation cloud, will be discussed in  \cite{guth_effect_2024}. We note though, that while profile gradients induce a rocket force outward in the minor radius, thus counteracting penetration of the pellet (in all toroidal magnetic confinement schemes), shielding induced asymmetries tend to accelerate pellets against the field line curvature vector. In tokamaks this results in motion towards the low magnetic field side, making the shielding contribution beneficial for high field side injection, while in stellarator magnetic confinement configurations the picture is more complex, owing to the strong toroidal variation of the curvature vector. 

In summary, we have formulated a closed expression for the rocket acceleration experienced by a pellet of frozen material travelling through a hot MCF plasma. We identify the pressure asymmetry as the dominant contribution to the force on the ablating pellet, and provide a simple expression for it that is linear in both the asymmetry of the heat flux and that of the representative electron energy reaching the neutral cloud. 
While plasmoid shielding effects have been expected to play a significant role, retaining only the asymmetry produced by the background plasma gradients leads to rocket acceleration predictions comparable with experimental measurements. 
Projections for ITER indicate that H-mode plasmas could slow down pellets injected from the low-field side enough to significantly reduce the fueling efficiency.
Taking into account the pellet rocket effect will therefore be crucial to ensure accurate predictions of the pellet ablation profiles in ITER.

\emph{Acknowledgements} --
The authors are grateful to E.~Nardon and P.~Aleynikov for fruitful discussions. The work was supported by the Swedish Research Council (Dnr.~2022-02862 and 2021-03943) and part-funded by the EPSRC Energy Programme [grant number EP/W006839/1]. The work has been partly carried out within the framework of the EUROfusion Consortium, funded by the European Union via the Euratom Research and Training Programme (Grant Agreement No 101052200 — EUROfusion). Views and opinions expressed are however those of the authors only and do not necessarily reflect those of the European Union or the European Commission. Neither the European Union nor the European Commission can be held responsible for them. 

\bibliography{references_from_Zotero.bib,references_manually_added.bib}

\end{document}